\documentclass[notitlepage,showpacs,superscriptaddress,nobalancelastpage,aps,prb,twocolumn,floatfix,10pt]{revtex4-2}

\usepackage[english]{babel}
\RequirePackage[T1]{fontenc}
\RequirePackage{times}

\usepackage{siunitx}
\usepackage[italicdiff]{physics}
\usepackage{amsfonts}
\usepackage{amsmath}
\usepackage{amssymb}
\usepackage{graphicx}
\usepackage{subfigure}
\usepackage[dvipsnames]{xcolor}
\usepackage{appendix}
\usepackage{hyperref}
\usepackage{my_symbols}
\usepackage{CJK}
\usepackage{bm}
\usepackage{tikz}
\usepackage{booktabs}
\usepackage{multirow}
\usepackage{pifont}
\usepackage{setspace}

\begin{document}
\begin{CJK*}{UTF8}{gbsn} 
	\title{Anatomy of spin wave driven magnetic texture motion via magnonic torques}
	\author{Hanxu Ai (艾寒旭)}
	\affiliation{Center for Joint Quantum Studies and Department of Physics, School of Science, Tianjin University, Tianjin 300354, China}
	\author{Jin Lan (兰金)}
	\email[Corresponding author:~]{lanjin@tju.edu.cn}
	\affiliation{Center for Joint Quantum Studies and Department of Physics, School of Science, Tianjin University, Tianjin 300354, China}
	\affiliation{Tianjin Key Laboratory of Low Dimensional Materials Physics and Preparing Technology, Tianjin University, Tianjin 300354, China}

	\begin{abstract}
		The interplay between spin wave and magnetic texture represents the information exchange between the fast and slow dynamical parts of magnetic systems.
        Here we formulate a set of magnonic torques acting on background magnetic texture, by extracting time-invariant information from the fast precessing spin waves.
		Under the frame of magnonic torques, we use theoretical formulations  and micromagnetic simulations to investigate the spin wave driven domain wall motion in two typical symmetry-breaking situations:
		the rotational symmetry broken by the Dzyaloshinkii-Moriya interaction, 
		and the translational symmetry broken by magnetic damping.
		The torque-based  microscopic analyses provide compact yet quantitative tools to reinterpret the magnetic texture dynamics induced by spin wave, beyond the conventional framework of global momentum conservation.
	\end{abstract}
	\maketitle
\end{CJK*}

\section{Introduction}

Spin wave, the propagating disturbance of ordered magnetizations, is one of the basic excitations in magnetic systems.
As an alternative angular momentum carrier besides the spin-polarized conduction electron \cite{kajiwara_transmission_2010}, spin wave possesses advantages of high energy efficiency, the ability of propagating in magnetic insulators \cite{kajiwara_transmission_2010, cornelissen_long-distance_2015,lebrun_tunable_2018} and the integrability with current magnetic storage techniques \cite{han_mutual_2019}.
With above merits and the rapidly developing experimental techniques \cite{nambu_observation_2020,wang_noninvasive_2022}, spin wave is exposed to growing interests from both theoretical and experimental sides \cite{chumak_magnon_2015,barman_2021_2021}.

Due to intrinsic nonlinearity of various magnetic interactions, the propagation of spin wave is expected to induce rich dynamics of underlying magnetic texture.
Using spin wave to harness the magnetic texture, as well as the converse manipulation of spin wave via magnetic texture, offer an integrated scheme toward constructing purely magnetic computing devices \cite{lan_spin-wave_2015, yu_magnetic_2020}.
Despite the urgent demand, transparent understandings on the spin wave driving scenarios are impeded the complexities brought by the fast precession of spin wave, as well as the nonuniform magnetization distribution of magnetic texture.
To overcome this obstacle, a powerful approach is exploiting momentum conservation laws that focus on global momentum transfer of linear momentum \cite{wang_magnondriven_2015,yu_polarization-selective_2018,rodrigues_spin-wave_2021}, spin angular momentum  \cite{yan_all-magnonic_2011,hinzke_domain_2011,mochizuki_thermally_2014} and orbital angular momentum \cite{jia_twisted_2019,jiang_twisted_2020,lee_orbital_2021}  as well as their combinations \cite{tveten_antiferromagnetic_2014,kim_propulsion_2014, oh_bidirectional_2019,yu_magnetic_2021}.
However, requirements of symmetry-preservation impose stringent limitations on the applicable range of above momentum transfer scenarios to realistic magnetic systems.

To escape the symmetry limitations and gain more insights into the actions of spin wave, microscopic theories focusing on the interaction details are called for, where a useful strategy is to make the interaction simple in either space or time side.
Spatially local interaction can be realized by constructing spin wave packet, and transforming spin wave packet and magnetic texture into two particle-like objects \cite{daniels_topological_2019,lan_skew_2021,jin_magnon-driven_2021,lan_spin_2022}. 
On the other hand, time invariant interaction can be realized by extracting the magnonic torques from the spin wave, and discard irrelevant information in fast oscillations.
The magnonic torques,  although frequently in an incomplete form, have been derived and employed to explore specific magnetic dynamics in several literatures \cite{kovalev_thermomagnonic_2012,kovalev_skyrmionic_2014,manchon_magnon-mediated_2014,kim_landau-lifshitz_2015}.
Nevertheless, insufficient cross-checking with micromagnetic simulations, and the lack of explicit connections with other theoretical models, especially with the conventional momentum transfer models, hinder the full application of the magnonic torques or their variants.

In this work, we systematically recast actions of spin wave on magnetic texture to a full set of magnonic torques, and further establish the generalized Thiele equation for magnetic texture dynamics via these torques.
To demonstrate the applicability of these microscopic torques, we investigate the spin wave driven domain wall motion in the presence of Dzyaloshinkii-Moriya interaction (DMI) and magnetic damping, for which both rotational and translational symmetries are broken.
By capturing the domain wall dynamics in such a typical symmetry-breaking system to the frame of magnonic torques, we partition the overall contribution to multiple aspects of spin wave. 
The detailed analyses enabled by magnonic torques, offer insights in developing purely magnetic information processing technologies.

The rest of paper is organized as follows.
In Sec. II, we systematically derive magnonic torques from fast oscillating spin waves, and then formulate a generalized Thiele equation to describe the magnetic texture dynamics induced by these magnonic torques.
With the aid of these magnonic torques, a hierarchy for various magnonic driving models is also proposed.
Using the simplified scenarios based on magnonic torques, the spin wave driven domain wall motion in presence of DMI and magnetic damping, as two typical situations of broken symmetry, are then thoroughly investigated in Sec. III. 
Comparisons between actions mediated by magnons and electrons, as well as a short conclusion, are given in Sec. IV.

\section{Magnonic torques and generalized Thiele equation}
\subsection{Basic model}
Consider a magnetic system with its magnetization direction denoted by a unit vector $\mb$, then the magnetic dynamics is governed by the Landau-Lifshitz-Gilbert (LLG) equation
\begin{align}
	\label{eqn:LLG}
	\dot{\mb} = -\gamma \mb\times \bh +\alpha \mb \times \dot{\mb},
\end{align}
where $\dot{\mb}\equiv\partial_t\mb$,  $\gamma$ is the gyromagnetic constant, and $\alpha$ is the Gilbert damping constant.
Here $\bh=-(1/\mu_0 M_s)\delta U/\delta \mb$ is the effective magnetic field acting on the magnetization $\mb$, where $\mu_0$ is the vacuum permeability, $M_s$ is the saturation magnetization, and $U=\int u d \cV$ is the magnetic energy with $\cV$ the magnetic volume.
The magnetic energy density $u$ takes a typical form of uniaxial magnets,
\begin{align}
	u=  \frac{\mu_0 M_s}{2} \qty[A (\nabla \mb)^2 +K(1-m_z^2)+ D \mb\cdot \qty((\bp\times \nabla)\times \mb)],
\end{align}
where $A$ is the exchange constant, $K$ is the easy-axis anisotropy along $\hbz$, and $D$ is the strength of the (interfacial-type) DMI with $\bp$ denoting the symmetry breaking direction.
The dipolar interaction is not considered explicitly, since for exchange type spin wave of interest in this work, its main role is to renormalize the anisotropy \cite{gurevich_Magnetization_1990}.

We suppose that the DMI is moderate with $D<D_c\equiv4\sqrt{AK}/\pi$, hence the ground state is always the homogeneous magnetization $\mb=\pm \hbz$ with $u=0$ \cite{rohart_skyrmion_2013}.
Depending on the exact combinations of magnetic parameters $A$, $K$ and $D$, various magnetic textures such as domain walls \cite{yu_magnetic_2021}, skyrmions \cite{back_2020_2020}, bobbers \cite{rybakov_new_2015} and hopfions \cite{gobel_skyrmions_2021} may be stabilized with finite energy density $u>0$.

When temporal evolution is involved, the total magnetization naturally divides into the slowly varying magnetic texture $\mb_0$ and the fast oscillating spin wave $\mb'$. 
Due to unity constraint $|\mb|=1$,  the transverse condition $\mb' \cdot \mb_0=0$ is satisfied everywhere.
Hence, we may define the spherical coordinates $\hbe_{r,\theta,\phi}$ adhered to texture magnetization $\hbe_r\equiv \mb_0$, and describe the spin wave as $\mb'=m_\theta\hbe_\theta+m_\phi\hbe_\phi$.

\subsection{Spin wave density and flux}

Before entering into explicit formulations regarding the actions on magnetic texture, we make following observations for spin wave:
i) In uniaxial ferromagnets of interest here, spin wave can always be treated as right-circularly polarized, even when travelling upon inhomogeneous magnetic texture.
ii) Because of much faster oscillation of spin wave, pre-processing spin wave information in each oscillation period does not affect the remarkably slower dynamic of magnetic texture.

Based on above observations, we proceed to extract time-invariant information from fast oscillating spin wave.
Apparently, $1$st order correlations vanish, $\langle m_{\theta/\phi} \rangle =0$, 
where $\langle \dots \rangle$ denotes the time-averaging evaluation.
Meanwhile, $2$nd order spin wave correlations yield
\begin{subequations}
	\label{eqn:sw_2nd_avg}
	\begin{align}
		\langle m_\theta^2 \rangle =            & \langle m_\phi^2 \rangle=\rho,                                 \\
		\langle m_\phi \nabla m_\theta \rangle= & -\langle m_\theta \nabla m_\phi \rangle=\frac{\bj}{2\gamma A},
	\end{align}
\end{subequations}
where $\rho=\langle \mb'\cdot \mb'\rangle /2$ and  $\bj= \gamma A \langle \nabla\mb'\times \mb'\rangle\cdot \mb_0$ are the spin wave density and flux, respectively.
Besides, other correlations include $\langle m_\theta m_\phi \rangle=0$ and $\langle m_\theta \nabla m_\theta\rangle=\langle m_\phi \nabla m_\phi \rangle =\nabla \rho/2$.

\subsection{Magnonic torques}

By partitioning into magnetic texture $\mb_0$ and spin wave $\mb'$, the total magnetization reads $\mb=\sqrt{1-\mb'\cdot\mb'}\mb_0+\mb'$, where the magnitude of the texture magnetization $\mb_0$ is subject to a reduction factor $\sqrt{1-\mb'\cdot\mb'}$ as enforced by unity constraint $|\mb|=1$. 
In accordance with above partition scheme, the LLG equation \eqref{eqn:LLG} is recast to
\begin{align}
	\label{eqn:LLG_tau}
	\dot{\mb}_0-\alpha \mb_0\times \dot{\mb}_0=\bm{\tau}_0+\bm{\tau},
\end{align}
where $\bm{\tau}_0=\gamma \bh(\mb_0)\times\mb_0 $ is the torque caused by texture gyration and distortion, and $\bm{\tau}$ is the magnonic torque mediated by spin wave $\mb'$. Performing time-averaging and utilizing the spin wave correlations in \Eq{eqn:sw_2nd_avg}, the magnonic torque $\bm{\tau}$ is then described by (see Appendix \ref{sec:ap_mag_tq} for detailed derivations)
\begin{align}
	\label{eqn:tau_sw}
	\bm{\tau} = & \bm{\tau}_\ssf{STT}+\bm{\tau}_\ssf{DM}+\bm{\tau}_\ssf{GL}+\bm{\tau}_\ssf{PL}  \nonumber \\
	= & (\bj \cdot\nabla)\mb_0- \frac{D}{2A}   (\bp\times \bj) \times \mb_0\nonumber               \\
	& - 2\rho\gamma \bh(\mb_0)\times\mb_0
	- \gamma A (\nabla \rho  \cdot \nabla)  \mb_0\times \mb_0,
\end{align}
where the $1$st and $2$nd terms are the spin-transfer torque (STT) and Dzyaloshinkii-Moriya (DM) torque mediated by the spin wave flux $\bj$, and $3$th and $4$th terms are gravity-like (GL) and pressure-like (PL) torques mediated by the spin wave density $\rho$, respectively.

The spin-transfer torque $\bm{\tau}_\ssf{STT}$ is due to the tracking of spin wave precession to the inhomogeneous texture magnetization \cite{kovalev_thermomagnonic_2012,kovalev_skyrmionic_2014}, similar to the process of electron spin tracking the background magnetization direction.
The DM torque  $\bm{\tau}_\ssf{DM}$ can be regarded as an extension of the spin transfer torque in the spirit of the chiral derivative \cite{kim_chirality_2013,kikuchi_dzyaloshinskii-moriya_2016}: $\partial_\beta \to \partial_\beta -(D/2A)(\bp\times\hbe_\beta)\times$, and takes analogy to the Rashba torque in electronic case \cite{manchon_magnon-mediated_2014}.
The gravity-like torque $\bm{\tau}_\ssf{GL}$ is caused by the rescaling of the texture magnetization $(1-\rho)\mb_0$
in the presence of spin wave $\mb'$ \cite{lan_skew_2021}.
The pressure-like torque $\bm{\tau}_\ssf{PL}$ is caused by the inhomogeneous spin wave distribution \cite{kim_landau-lifshitz_2015}, 
and is intimately related to the entropic torque \cite{hinzke_domain_2011, schlickeiser_role_2014}.

In a homogeneous domain with $\mb_0=\pm \hbz$, only the DM torque in \Eq{eqn:tau_sw} survives, for a proper combination of the spin wave current $\bj$, symmetry breaking direction $\bp$ and the magnetization $\mb_0$.
The specific form of DM torque indicates that the spin wave flux $\bj$ generates an effective magnetic field $\bh_D=-(D/2\gamma A)(\bp\times \bj)$ acting on the static magnetization $\mb_0$, and further leads to a small magnetization titling \cite{manchon_magnon-mediated_2014} (see also Appendix \ref{sec:ap_mag_dm}).
For bulk-type DMI instead of the interfacial-type here, the DM torque is simply replaced by $\bm{\tau}_\ssf{DM}=(D/2A)\bj\times \mb_0$. 

When the inhomogeneous distribution of spin wave density is purely induced by magnetic damping, following relation is satisfied  $\nabla\rho \approx -(\alpha d/2)(\bj/\gamma A) $, where $d$ is the system dimension \cite{kim_landau-lifshitz_2015}. 
For this purely dissipative case, the pressure-like torque is alternatively written as \cite{kovalev_skyrmionic_2014,kim_landau-lifshitz_2015}
\begin{align}
	\label{eqn:tau_PL_damping}
	\bm{\tau}_\ssf{PL}=\frac{\alpha d}{2} (\bj\cdot \nabla )\mb_0\times \mb_0=\frac{\alpha d}{2}\bm{\tau}_\ssf{STT}\times \mb_0,
\end{align}
which serves as dissipative (or non-adiabatic) correction \cite{zhang_roles_2004, thiaville_micromagnetic_2005} to the (adiabatic) spin-transfer torque with a factor $\alpha d/2$. 
The non-adiabatic nature of $\bm{\tau}_\ssf{PL}$ in \Eq{eqn:tau_PL_damping} originates from the slight mis-tracking of spin wave precession to the background magnetization variation, which we attribute to the response lag caused by magnetic damping.

\subsection{Generalized Thiele equation}

As seen in Eqs. \eqref{eqn:LLG_tau}\eqref{eqn:tau_sw}, trimming the actions of rapidly oscillating spin waves into several magnonic torques markedly  reduces the complexity of magnetic dynamics.
Since magnetic texture tends to maintain its shape, further complexity reduction can be achieved by capturing the  magnetic texture dynamics via the evolution of collective coordinates $\qty{X_\mu}$ with $\mu=1,2,3,\dots$, i.e., $\mb_0(t)\equiv\mb_0[\qty{X_\mu(t)}]$.
Specifically, left multiplying $\partial_\mu \mb_0\times\mb_0 $ to \Eq{eqn:LLG_tau} and integrating in the whole magnetic volume $\cV$ yields the generalized Thiele equation \cite{thiele_steady-state_1973,tretiakov_dynamics_2008},
\begin{align}
	\label{eqn:force_tx_sw}
	&B^0_{\mu\nu}  \dot{X}_\nu +E^0_\mu- \alpha \Lambda^0_{\mu\nu} \dot{X}_\nu \nonumber \\
	=  &\int \qty[   (b^0_{\mu\beta}+ b^D_{ \mu \beta} )j_\beta
		+2   e^0_\mu  \rho -    \gamma A\lambda^0_{\mu\beta}  \partial_\beta \rho]d\cV,
\end{align}
where $\mu$ and $\nu$ are indices reserved for collective coordinates of magnetic texture, $\beta$ is for Cartesian coordinates, and Einstein summation rule on repeated indices is implied.
Here $b^0_{\mu \beta }= \mb_0 \cdot(\partial_\mu \mb_0 \times \partial_\beta \mb_0)$ and $b^D_{\mu \beta }= (D/2A)(\bp\times\hbe_\beta)\cdot \partial_\mu \mb_0$
are the fictitious magnetic fields caused by the magnetic topology and DMI \cite{van_hoogdalem_magnetic_2013,kim_tunable_2019,lan_skew_2021},
$e^0_\mu=-(\gamma/\mu_0 M_s)\partial_\mu u_0$ 
is the fictitious electric field caused by the inhomogeneous magnetic energy density \cite{lan_skew_2021,lan_geometric_2021}, $\lambda^0_{\mu\beta}=\partial_\mu \mb_0 \cdot \partial_\beta \mb_0$ is the adhesion field characterizing the `roughness' of the nonuniform magnetizations.
The global quantities denoted in upper case follows the definition of above local quantities in lower case, i.e., $B^0_{\mu\nu} =\int b^0_{\mu\nu} d\cV$, $E^0_\mu=\int e^0_\mu d\cV$, and $\Lambda^0_{\mu\nu}=\int \lambda^0_{\mu\nu} d \cV$.
Alternatively for magnetic texture, $B_{\mu\nu}^0$ is the gyroscopic coefficient correlated with the topology charge, $E^0_\mu=-(\gamma/\mu_0M_s)\partial_\mu U_0$ is the restoration force, and $\alpha \Lambda^0_{\mu\nu}$ is the viscous coefficient.

In \Eq{eqn:force_tx_sw}, all magnonic torques are transformed to forces mediated by the fictitious fields generated by the magnetic texture (see Appendix \ref{sec:ap_mag_forces} for detailed discussions):
The spin-transfer torque $\bm\tau_\ssf{STT}$ and the DM torque $\bm\tau_\ssf{DM}$ to Amp\`ere forces, the gravity-like torque $\bm\tau_\ssf{GL}$ to  electrostatic force, and the pressure-like torque $\bm\tau_\ssf{PL}$ is converted to adhesion force. Apparently, the two sides of \Eq{eqn:force_tx_sw} share the same fields $b^0$, $e^0$ and $\lambda^0$, reminding that magnetic texture and spin wave are under the same magnetic environment. 
The only unpaired force is due to the additional field $b^D$, which highlights the unique role of DMI in shaping magnetic dynamics.
Indeed, DMI is a key ingredient that introduces chirality to magnetic texture \cite{yang_firstprinciples_2022}, and non-reciprocity to spin wave \cite{moon_spin-wave_2013,cortes-ortuno_influence_2013}.

By decomposing continuous spin wave into discrete spin wave packets, Amp\`ere forces exerted by flux $j$ in \Eq{eqn:force_tx_sw} divides into Lorentz forces exerted by moving charges.
Consequently, the interplay between spin wave packet and magnetic texture mimics the collision between two particle-like objects,  as established in Ref. \onlinecite{lan_skew_2021}.

\subsection{Hierarchy of magnonic driving models}

\begin{table*}[tb]
	\centering
	\caption{Hierarchy of magnonic driving models.}
	\label{tab:hierarchy}
	\tabcolsep=0.35cm
	\begin{spacing}{1.2}
	\begin{tabular}{*{8}{c}}
		\midrule
		\midrule

		\multirow{3.5}*{Level} & \multirow{3.5}*{Core approach}&\multirow{3.5}*{Reference} & \multicolumn{5}{c}{Complexity} \\

		\cmidrule(lr){4-8}

		& & & \multicolumn{2}{c}{Spin wave} & \multicolumn{2}{c}{Magnetic texture} & \multirow{2.2}*{Interaction}\\

		\cmidrule(lr){4-5}\cmidrule(lr){6-7}
		
		& & & Time & Space & Time & Space & \\

		\midrule

		Wave  &\textemdash & LLG equation \eqref{eqn:LLG} & \ding{51}  & \ding{51} &\textemdash&\ding{51}& \ding{51} \\

		Torque  & Time averaging & \Eq{eqn:tau_sw}, Refs. \onlinecite{kovalev_thermomagnonic_2012,kovalev_skyrmionic_2014,manchon_magnon-mediated_2014,kim_landau-lifshitz_2015} & \ding{55} & \ding{51} &\textemdash& \ding{51} & \ding{51} \\

		Force & Thiele equation & Eq. \eqref{eqn:force_tx_sw}, Refs. \onlinecite{thiele_steady-state_1973,tretiakov_dynamics_2008} & \ding{55} & \ding{51} &\textemdash& \ding{55} & \ding{51} \\

		Particle & Spin wave packet & 
		Refs. \onlinecite{daniels_topological_2019,lan_skew_2021,jin_magnon-driven_2021,lan_spin_2022}  & \ding{55}  & \ding{55} &\textemdash& \ding{55} & \ding{51} \\
		
		Momentum & Noether theorem & Refs. \onlinecite{yan_all-magnonic_2011,wang_magnondriven_2015,yu_magnetic_2021}    & \textemdash  & \textemdash &\textemdash& \ding{55} & \ding{55} \\

		\midrule
		\midrule
	\end{tabular}
\end{spacing}
\end{table*}

From preceding discussions,  a hierarchy of magnonic driving models can be constructed by making more and more simplifications in each stage, as list in Table \ref{tab:hierarchy}.
Apparently, above $5$ models are lying at different levels of complexities, thus have own advantages and limitations in handling specific problems.

\begin{figure*}[tb]
	\centering
	{\includegraphics[width=0.98 \textwidth,  trim=10 0 10 20, clip]{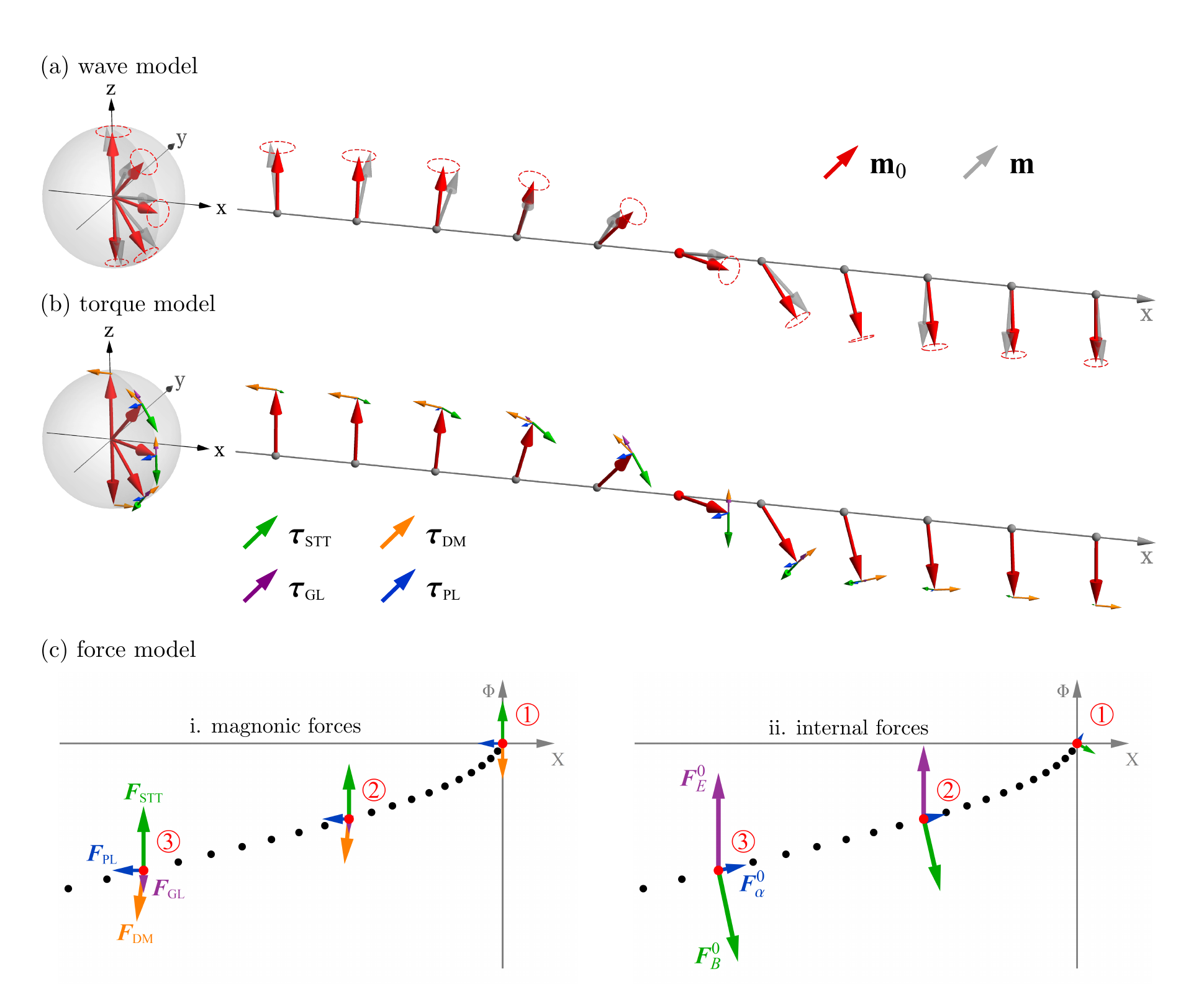}}
	\caption{
	{\bf Schematics of the domain wall dynamics induced by (a) spin wave, (b) the equivalent magnonic torques and (c) magnonic forces.
	}
	In (a)(b), the red arrows depict the domain wall magnetizations, and all magnetizations are united in a magnetic Bloch sphere.
	For (a), the red circles represent the precessing spin waves; and for (b), the green/orange/purple/blue arrows represent the corresponding magnonic torques. 
	In (c), the arrows depict the magnonic/internal forces (see Appendix \ref{sec:ap_mag_forces} for definitions) at the left/right panel at $3$ selected time. 
	The black dots plot the time evolution of domain wall in parametric space $\qty{X,\Phi}$, and the red dots depict the $3$ observation points.
	\label{fig:dwm_DM_sch}
	}
\end{figure*}

\section{Domain wall motion induced by magnonic torques}

In the following, we scrutinize the torque model established in preceding section by investigating the spin wave driven domain wall motion in a magnetic wire, where both DMI and magnetic damping are present. 
Specifically,  the DMI breaks the rotational symmetry, and the magnetic damping breaks the translational symmetry in both space and time.
While the momentum model hinging on symmetry-preservation is no longer applicable, the microscopic torque model naturally fits into such investigations.

\subsection{Theoretical formulations}

Consider a magnetic wire extending in $x$-axis, where the easy axis of the anisotropy and the symmetry-breaking direction of DMI both lie in $z$-axis.
The ground states are magnetic domains with $\mb_0=\pm \hbz$, and the domain wall lies in between two domains takes N\'{e}el type as enforced by DMI, as seen in Fig. \ref{fig:dwm_DM_sch}.
More explicitly, the domain wall takes the Walker profile with $\theta_0=2\arctan[\exp(x/W)]$ and $\phi_0=0$, where $\theta_0$ and $\phi_0$ are the polar and azimuthal angles of $\mb_0$ about $\hbz$, and $W=\sqrt{A/K}$ is the characteristic width.
The domain wall energy is $u_0= \mu_0 M_s K \sech^2(x/W)$, which is contributed by the exchange coupling and the anisotropy half by half.
The slow variation of the domain wall can be described by
\begin{align}
	\label{eqn:dmw_xphi}
	\theta_0(t)=2 \arctan\qty[\exp(\frac{x-X(t)}{W})], \quad \phi_0(t)=\Phi(t),
\end{align}
where the central position $X$ and rotation angle $\Phi$ constitute the minimal set of collective coordinates of domain wall.

Denoting the spin wave by complex field $\psi=m_\theta-im_\phi$, the spin wave dynamics is recast from LLG equation \eqref{eqn:LLG} to a Schr\"odinger-like equation,
\begin{align}
	\label{eqn:sw_schrodinger}
	\frac{i-\alpha}{\gamma}\dot{\psi}  = \qty(-A\partial_x^2+K-\frac{2}{\mu_0 M_s}u_0+\frac{D }{2W}m_0^x)\psi,
\end{align}
where two effective potentials arise due to inhomogeneous domain wall profile and DMI, respectively.
Since the reflection by the P\"oschl-Teller type potential well $-(2/\mu_0 M_s)u_0$ is always absent, and the scattering by the additional potential $(D/2W)m_0^x$ is negligible except for extremely low frequency, here we suppose perfect spin wave transmission for all cases.
Consequently, the spin wave travelling upon the domain wall governed by \Eq{eqn:sw_schrodinger}  simply takes a decaying plane wave form,
\begin{align}
	\label{eqn:wave_sw_1d}
	\psi=c\exp(ikx-\kappa x -i\omega t),
\end{align}
where $c$ is the amplitude, $k$ and $\kappa$ are the real and imaginary parts of the wavevector, and $\omega$ is the angular frequency.
The spin wave dispersion is $(1+i\alpha)\omega = \gamma [A(k+i\kappa)^2+K]$, and the group velocity is $v\equiv\partial_k \omega =2\gamma A k$.

From above spin wave information in \Eq{eqn:wave_sw_1d}, the spin wave density is  $\rho=(c^2/2) \exp(-2\kappa x)$, the spin wave flux is $j=2 \gamma\rho A  k=\rho v $, and the density gradient is $\partial_x\rho=-2\rho \kappa$.
Therefore, the magnonic torques in \Eq{eqn:tau_sw}  are explicitly given by
\begin{align}
	\label{eqn:tau_sw_1d}
	\bm{\tau}= & \bm{\tau}_\ssf{STT}+\bm{\tau}_\ssf{DM}+\bm{\tau}_\ssf{GL}+\bm{\tau}_\ssf{PL}  \nonumber \\
	=& \rho v\qty(\partial_x \mb_0 +\frac{D}{2A} \mb_0\times \hby) \nonumber        \\
	+           & 2\rho\qty(  \mb_0\times \gamma\bh_0- \kappa \gamma A \mb_0\times \partial_x  \mb_0),
\end{align}
where all torques are proportional to the spin wave density $\rho$, i.e., lying in second order to spin wave.

With magnonic torques in \Eq{eqn:tau_sw_1d}, the domain wall dynamics with collective coordinates $X$ and $\Phi$ is recast from \Eq{eqn:force_tx_sw} to
\begin{subequations}
	\label{eqn:dw_XP}
	\begin{align}
		\label{eqn:dw_XP_X}
		\frac{\dot X}{W}-\alpha \dot\Phi = &
		-\frac{\rho_0 v}{W}\qty(1-\tilde{D}\cos\Phi)+ \qty(1-2\rho_0)\frac{v}{W} \frac{\tilde{D} \sin\Phi}{2kW}, \\
		\label{eqn:dw_XP_P}
		\dot\Phi+\frac{\alpha \dot X}{W}=  & 2 \kappa \rho_0 v 
		\qty(  \tilde{D} \sin\Phi- \frac{\varepsilon}{2kW}),
	\end{align}
\end{subequations}
where $\rho_0=\rho_s \exp[- 2\kappa (X-x_s)]$ is the spin wave density at the domain wall center, and $x_s$ and $\rho_s$ are the position and spin wave density of spin wave source, respectively.
Here, a correction factor $\varepsilon=3/2$ is artificially introduced to take account of the domain wall dynamics beyond the parametric space $\qty{X,\Phi}$ (see Appendix \ref{sec:ap_mag_dm} for details).
Noting that $\kappa\approx \alpha \omega/v$ for small damping $\alpha\ll 1$,  the leading order of the domain wall motion in $\dot{X}$ (rotation in $\dot{\Phi}$) then lies at the $0$th ($1$st) order of the damping constant $\alpha$. 
In this regard, the domain wall motion in $\dot{X}$ is mainly contributed by spin transfer torque $\bm{\tau}_\ssf{STT}$, DM torque $\bm{\tau}_\ssf{DM}$ and internal torque $\bm{\tau}_0$, while the contribution of $\bm{\tau}_\ssf{GL}$ is negligible due to smallness of rotation angle $\Phi\ll 1$ and spin wave density $\rho\ll 1$. In the meantime, beside the contribution of DM torque $\bm{\tau}_\ssf{DM}$ and pressure-like torque $\bm{\tau}_\ssf{PL}$ induced by the spin wave attenuation, the hybridization term between $X$ and $\Phi$ also manifests as a crucial ingredient in rotating domain wall.
Alternatively, \Eq{eqn:dw_XP} can be interpreted as a balance between external magnonic forces and internal texture forces, as elaborated in Appendix \ref{sec:ap_mag_forces}.

The spin wave in \Eq{eqn:wave_sw_1d} together with the corresponding magnonic torques formulated in \Eq{eqn:tau_sw_1d} and the magnonic forces formulated in \Eq{eqn:force_sw_1d} , are schematically depicted in Fig. \ref{fig:dwm_DM_sch}.
Apparently, by converting spin wave to magnonic torques and further to magnonic forces, the driving scenarios become increasingly transparent.
For torque model in Fig. \ref{fig:dwm_DM_sch}(b), both the spin transfer torque $\bm{\tau}_\ssf{STT}$ and gravity-like torque $\bm{\tau}_\ssf{GL}$ tend to move the domain wall, while the pressure-like torque $\bm{\tau}_\ssf{PL}$ tends to rotate the domain wall.
Meanwhile, the DM torque $\bm{\tau}_\ssf{DM}$ competes with all these torques in  both domain wall motion and rotation.
Above relations between magnonic torques are corroborated by the force model in Fig. \ref{fig:dwm_DM_sch}(c), where the participation of DMI in both dynamics of $X$ and $\Phi$ is unambiguously indicated by the tilting of $\bF_\ssf{DM}$. 
In addition, the tilting of both $\bF^0_{B}$ and $\bF^0_\alpha$ also highlight the dynamics hybridization between $X$ and $\Phi$, due to gyroscopic nature of ferromagnetic dynamics.

\begin{figure*}[tb]
	\centering
	{\includegraphics[width=0.95 \textwidth]{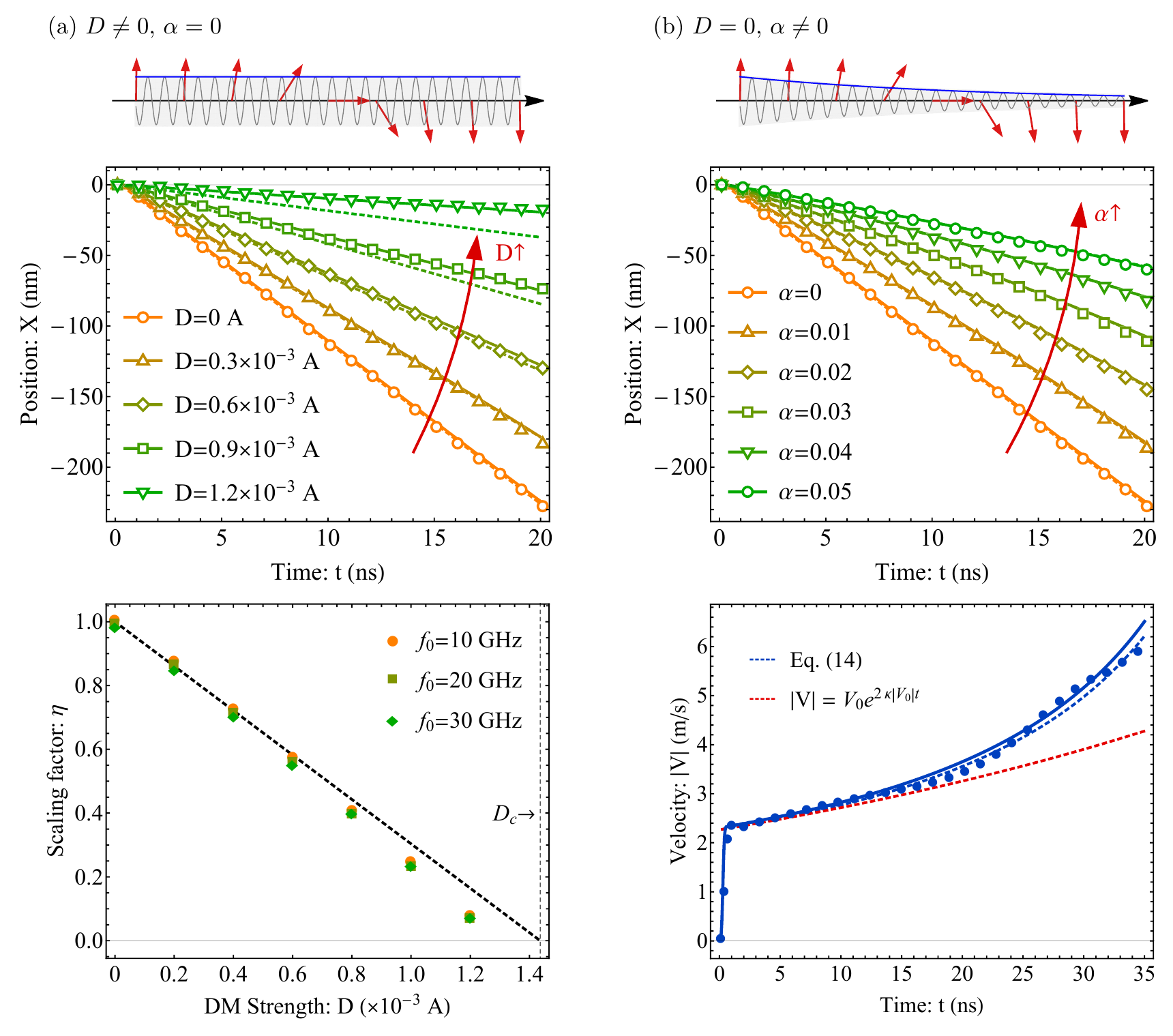} }
	\caption{{\bf Spin wave driven domain wall motion under  (a) $D\neq0$, $\alpha=0$  and   (b) $D=0$, $\alpha\neq 0$.}
		In (a), the upper panel plots the time evolution of domain wall position under a set of different DM strength $D$, the lower panel plots the scaling factor $\eta$ as function of DM strength.
		In (b), the upper panel plots the time evolution of domain wall position under a set of magnetic damping $\alpha$, the lower panel plots the domain wall velocity $V$ for a relatively long time.
		For all plots, the dots/solid lines are extracted from  micromagnetic simulations based on original LLG equation \eqref{eqn:LLG} and the magnonic torques  \eqref{eqn:tau_sw_1d}, and the dashed lines are based on  \Eq{eqn:dw_XP}.
		The spin wave is excited at $x_0=\SI{-300}{nm}$ with excitation frequency $f=\SI{10}{GHz}$ and spin wave density $\rho_0=\SI{0.02}{}$.
		\label{fig:dwm_DM_alphaD0}
	}
\end{figure*}

\subsection{Numerical results}

To quantitatively investigate the domain wall dynamics, we carry out three types of numerical calculations in parallel: 
i) Wave-based analysis, by performing micromagnetic simulations upon the original LLG equation \eqref{eqn:LLG} with spin wave included; 
ii) Torque-based analysis, by performing micromagnetic simulations upon the equivalent magnonic torques in \Eq{eqn:tau_sw_1d}; 
iii) Force-based analysis, by seeking numerical solutions to the collective coordinate equation \eqref{eqn:dw_XP}.
For micromagnetic simulations in wave- and torque-based analyses, a micromagnetic module developed by Yu et al. \cite{yu_micromagnetic_2021} is employed, where the LLG equation is transformed to a weak form, and solved by the generalized-alpha method.
The magnetic parameters are mainly based on the yttrium iron garnet (YIG) \cite{lan_spin-wave_2015,wu_curvilinear_2022}: the exchange coupling constant $A=\SI{3.28e-11}{A.m}$, the easy-axis anisotropy $K=\SI{3.88e4}{A.m^{-1}}$, the saturation magnetization $M_s=\SI{1.94e5}{A.m^{-1}}$, the gyromagnetic ratio $\gamma=\SI{2.21e5}{m.A^{-1}.s^{-1}}$ and the vacuum permeability $\mu_0=\SI{1.26e-6}{T.m.A^{-1}}$.

When damping is negligible, $\kappa=\alpha=0$, domain wall is subject to no rotation ($\Phi=0$) according to \Eq{eqn:dw_XP_P}, and the domain wall velocity is simply given by
\begin{align}
	\label{eqn:dmw_V_alpha0}
	V\equiv \dot{X}=- \eta \rho_s v,
\end{align}
where $\eta=1-\tilde{D}$. The scaling factor $\eta$ originates from the rotational symmetry breaking induced by DMI, by noting that the relation $V=-\rho_s v$ at $D=0$ is a manifestation of angular momentum conservation \cite{yan_all-magnonic_2011}.
The slowdown of the domain wall motion by DMI is clearly shown in Fig. \ref{fig:dwm_DM_alphaD0}(a), for a series of DM strength. Moreover, the factor $\eta$ extracted at different spin wave frequencies also coincides well with the theoretical expectation in \Eq{eqn:dmw_V_alpha0}.
Nevertheless, slight deviation from \Eq{eqn:dmw_V_alpha0} is spotted as $D$ approaches $D_c$ (or $\eta$ approach $0$), which we attribute to the increasing instability toward a spin spiral state.

When magnetic damping is finite $\alpha\neq 0$ albeit the DMI is absent $D=0$, the domain wall velocity is simply described by $V=-\rho_0 v$, according to \Eq{eqn:dw_XP_X}.
However, due to spin wave attenuation, the domain wall experiences larger magnonic torque as its approaches the spin wave source. 
As a result, domain wall velocity is modulated by $\dot{V}/V=\dot{\rho}_0/\rho_0 =-2\kappa V$, i.e., the domain wall is subject to an effective drag force with coefficient $2\kappa$.
Consequently, the domain wall velocity is explicitly described by \cite{lan_spin_2022}
\begin{align}
	\label{eqn:dmw_V_D0}
	V= -\frac{|V_0|}{1-2\kappa |V_0|t},
\end{align}
where $V_0 = -\rho_s v \exp[-2\kappa(X_0-x_s)]$ is the domain wall velocity at initial position $X_0$.
For a short time $t\ll t_0 \equiv 1/2\kappa |V_0|\approx \SI{55}{ns}$, the domain wall velocity is almost invariant as time, and decreases exponentially as damping constant $\alpha$ increases, as depicted in Fig. \ref{fig:dwm_DM_alphaD0}(b).
While for a long time $t\sim t_0$, the domain wall velocity increases reciprocally as formulated in \Eq{eqn:dmw_V_D0}, deviates from an empirically anticipated exponential law $V=-|V_0|\exp(2\kappa |V_0| t)$.

Although both the DMI and the damping slow down the domain wall motion driven by spin wave in Fig. \ref{fig:dwm_DM_alphaD0}, their roles are distinct in viewpoint of symmetry.
The main role of DMI is to generate a DM torque that counteracts the spin-transfer torque, i.e., the underlying mechanism is the rotational symmetry breaking.
In contrast, the role of damping is to attenuate the spin wave and thus decrease the spin wave density (flux) touching the domain wall, i.e.,  the underlying mechanism is translational symmetry breaking.

\begin{figure}[tb]
	\centering
	\includegraphics[width=0.48 \textwidth,trim=0 0 0 0,clip]{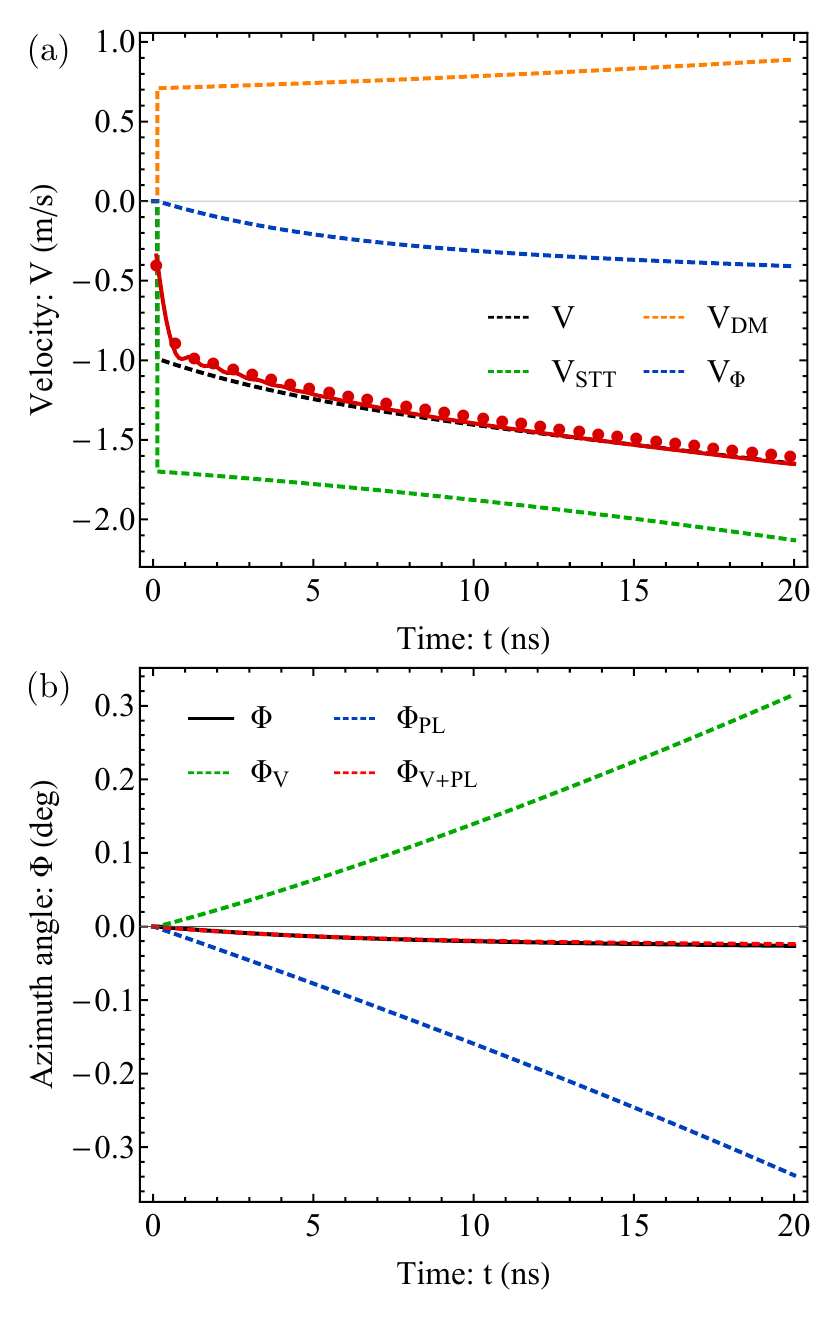}
	\caption{{\bf Time evolution of  (a) domain wall velocity $V$ and (b) angle $\Phi$   under $D=\SI{0.6e-3}{A}$ and $\alpha=\SI{0.05}{}$.}
	The dots are for wave-based analysis upon \Eq{eqn:LLG}, the solid lines are for torque-based analysis upon \Eq{eqn:tau_sw_1d}, and  dashed lines are for force-based analysis upon \Eq{eqn:dw_XP},  respectively.
		\label{fig:dwm_Dalpha_hybrid}
	}
\end{figure}

When sizable DMI and magnetic damping are both present with $D\neq 0$ and $\alpha \neq 0$, the evolution of domain wall velocity becomes more complicated, due to competitions and correlations between multiple torques in \Eq{eqn:tau_sw_1d}. Despite these complexities, the evolution of domain wall velocity is quantitatively reproduced by the both torque-based simulations and force-based analyses, as shown in Fig. \ref{fig:dwm_Dalpha_hybrid}(a).
Furthermore, according to \Eq{eqn:dw_XP}, the overall domain wall velocity naturally splits to three different parts: $V_\ssf{STT}$ by spin transfer torque, $V_\ssf{DM}$ from DM torque, $V_{\Phi}$ from the internal restoration torque.
A small contribution from GL torque also appears in \Eq{eqn:dw_XP_P}, but can be neglected due to the smallness of both rotation angle $\Phi$ and spin wave density $\rho$.
Among these contributions, $V_\ssf{STT}$ and $V_\ssf{DM}$ are much larger, and remain as the major part even after mutual cancellation; 
while $V_{\Phi}$ driven by finite angle $\Phi$ is also noteworthy, even though no spin wave reflection is included here.
To further investigate the driving mechanism, the rotation angle $\Phi$ is also divided into $2$ parts in Fig. \ref{fig:dwm_Dalpha_hybrid}(b):
$\Phi_{V}$ caused by dynamics hybridization between $X$ and $\Phi$, and $\Phi_\ssf{PL}$ caused by PL torque $\bm{\tau}_\ssf{PL}$. 
While $\Phi_{V}$ and $\Phi_\ssf{PL}$ always tend to cancel out with each other, the accumulation of the small mismatch gives rise to a small but gradually increasing rotation angle $\Phi$, and further modifies the domain wall velocity via $V_{\Phi}$ in Fig. \ref{fig:dwm_Dalpha_hybrid}(a).

\section{Discussions and conclusion}

The magnonic torque in \Eq{eqn:tau_sw} shares many similarities with its electronic counterpart, as manifested in the spin transfer torque and the DM (Rashba) torque, because both spin wave and spin-polarized conduction electron are angular momentum carriers \cite{kajiwara_transmission_2010}.
However, different from the electric nature of conduction electron, the magnetic nature of spin wave adds several new flavors to magnonic torques:
i) Unlike conduction electron, spin wave shares the same magnetic degree of freedom with the background magnetization, and thus the gravity-like and pressure-like magnonic torques additionally emerge.
ii) The intrinsic magnetic nature of spin wave implies its sensitivity to magnetic environment, therefore the spatial distribution of spin wave and further magnonic torques can be quite complicated due to magnetic scattering. 
In contrast, the electric current can be regarded as uniform in the length scale of magnetic texture, giving rise to much simpler form of its electronic counterpart.
iii) Due to relatively slow spin wave propagation, a finite time is required for magnonic torque to act on the whole magnetic texture, as demonstrated in Figs. \ref{fig:dwm_DM_alphaD0} and \ref{fig:dwm_Dalpha_hybrid};  
Given the extremely high propagation speed of electric current, such a response time is negligible.

In this work, the validity of magnonic torques rests with the circularly polarized form of spin wave, which is ensured when  exchange coupling is dominating. 
However, when hard-axis anisotropy is remarkable in biaxial ferromagnets, the spin wave is squeezed to elliptical form \cite{zou_tuning_2020}.
Furthermore, in antiferromagnets and ferrimagnets, spin wave is endowed with full polarization degree of freedom \cite{lan_antiferromagnetic_2017,yu_polarization-selective_2018,kim_tunable_2019,nambu_observation_2020,liu_switching_2022} and may take arbitrary polarization including all circular, linear and elliptical forms. 
To account for these complications, polarization and other relevant information beside the spin wave density and flux are necessary to fully describe the magnonic torques.
Above information would also add as ingredients to further distinguish the magnonic torque from its electronic counterpart.

In conclusion, by converting the action of spin wave to a set of magnonic torques, we established torque model to formulate the general dynamics of magnetic texture induced by spin wave.
Via the torque-based simulations and force-based analyses, the domain wall motion driven by spin wave is systematically investigated, even though the translational and rotational symmetries are broken by DMI and magnetic damping, respectively.
With unique features and transparent meanings of these magnonic torques, more delicate manipulations of magnetic texture using spin wave are envisioned.

\acknowledgments
J.L. is grateful to Jiang Xiao for insightful discussions.
This work is supported by National Natural Science Foundation of China (Grant No. 11904260) and Natural Science Foundation of Tianjin (Grant No. 20JCQNJC02020).

\appendix

\section{Derivation of magnonic torques}
\label{sec:ap_mag_tq}

When spin wave $\mb'$ travels upon magnetic texture $\mb_0$, the total magnetization subject to unity constraint $|\mb|=1$ is described by
\begin{align}
	\label{eqn:ap_mall}
	\mb=    & \sqrt{1-\mb'\cdot \mb'}\mb_0+\mb' \nonumber      \\
	\approx & (1-\frac{\mb'\cdot \mb'}{2})\mb_0+\mb' \nonumber \\
	=       & \mb_0+\mb'-\rho \mb_0,
\end{align}
which corresponds to $0$th, $1$st and $2$nd order terms in spin wave $\mb'$.
Meanwhile, the effective magnetic field is
\begin{align}
	\label{eqn:ap_heff}
	\bh(\mb)= A \nabla^2 \mb + K m_z\hbz - D(\bp \times \nabla)\times\mb,
\end{align}
which naturally splits to $3$ parts, $\bh(\mb)=\bh(\mb_0)+\bh(\mb')-\bh(\rho \mb_0)$, according to \Eq{eqn:ap_mall}.
Furthermore,  the effective field has the following property
\begin{align}
	\label{eqn:ap_hrhom}
	\bh(\rho\mb_0)-\rho\bh(\mb_0)
	\approx & 2A(\nabla\rho\cdot \nabla)\mb_0-D(\bp\times\nabla)\rho \times \mb_0,
\end{align}
which is caused by the fact $\partial_\beta(\rho \mb_0)-\rho \partial_\beta\mb_0=(\partial_\beta \rho) \mb_0$.

Inserting \Eq{eqn:ap_mall} to the LLG equation \eqref{eqn:LLG} yields
\begin{align}
	\label{eqn:ap_llg_sw}
	\dot\mb_0-\alpha\mb_0\times\dot{\mb}_0=- & \mb_0\times \gamma\bh(\mb_0)
	-\mb'\times \gamma\bh(\mb') \nonumber  \\
	+ & \rho \mb_0 \times \gamma\bh(\mb_0)+\mb_0\times \gamma\bh(\rho\mb_0),
\end{align}
where only $0$th and $2$nd order terms in spin wave $\mb'$ are kept, and the $1$st order terms are dropped since they always average out in time evolution.
On the right side of \Eq{eqn:ap_llg_sw}, the first term is $\bm{\tau}_0$ caused by the texture distortion and gyration,  and the remaining three items together constitute the magnonic torques $\bm{\tau}$.
Inserting \Eq{eqn:ap_heff} and \Eq{eqn:ap_hrhom} into \Eq{eqn:ap_llg_sw}, the magnonic torque is then described by
\begin{align}
	\label{eqn:ap_torque_ADK}
\bm{\tau}
	\approx   & 2\rho\mb_0\times \gamma \bh(\mb_0) \nonumber                      \\
	+& \mb_0\times \gamma \qty[2A(\nabla\rho\cdot \nabla)\mb_0-D(\bp\times\nabla)\rho \times \mb_0]\nonumber                     \\
	-  & \langle \mb'\times \gamma A \nabla^2\mb'\rangle+\langle \mb'\times \gamma \qty[D (\bp\times \nabla)\times \mb']  \rangle,
\end{align}
where the time-averaging $\langle \cdots \rangle$ is used to further remove redundant fluctuations.

In the $3$rd line of \Eq{eqn:ap_torque_ADK}, the torque related to the exchange coupling is transformed to
\begin{align}
	\label{eqn:ap_extq}
     &-\Avg{\mb'\times\gamma A \nabla^2\mb'}    \nonumber           \\
    =&  -\partial_\beta \Avg{\mb'\times\gamma A \partial_\beta \mb'}\nonumber           \\
	=       & \partial_\beta ( j_\beta \mb_0+\rho \partial_\beta \mb_0\times \mb_0)\nonumber \\
	\approx & (\bj\cdot\nabla)\mb_0+\gamma A(\nabla\rho\cdot \nabla)\mb_0\times\mb_0,
\end{align}
where following relation is employed,
\begin{align}
	\label{eqn:ap_extq_1}
	& \Avg{\mb'\times   \partial_\beta \mb'}\nonumber  \\
  = &- (\mb_0\cdot \Avg{\mb'\times   \partial_\beta \mb'})\mb_0+ \mb_0  \times(\Avg{\mb'\times   \partial_\beta \mb'}\times\mb_0)\nonumber \\
  = & -\frac{j_\beta}{\gamma A} \mb_0+\Avg{(\mb_0\cdot \partial_\beta \mb')(\mb'\times \mb_0)}\nonumber \\
  = & -\frac{j_\beta}{\gamma A} \mb_0-\Avg{(\partial_\beta  \mb_0\cdot \mb')(\mb'\times \mb_0)}\nonumber \\
  = &-\frac{j_\beta}{\gamma A} \mb_0-\rho \partial_\beta \mb_0\times \mb_0.
\end{align}
Similarly, the torque related to the DMI in  $3$rd line of \Eq{eqn:ap_torque_ADK} is transformed to
\begin{align}
	\label{eqn:ap_dmtq}
	&\Avg{\mb'\times \qty[\gamma D (\bp\times \nabla)\times \mb']}\nonumber \\
	= & \Avg{ \mb'\times \qty[\gamma D (\bp\times \hbe_\beta )\times \partial_\beta\mb']  } \nonumber \\
	= &	-  \Avg{\qty[\gamma D(\bp\times \hbe_\beta )\cdot\mb'] \partial_\beta \mb'}     + \gamma D (\bp\times \hbe_\beta ) \Avg{\mb'\cdot \partial_\beta\mb'} \nonumber   \\
	\approx &  -  \gamma D(\bp\times \hbe_\beta )\times \frac{ j_\beta}{2\gamma A}\mb_0 +\gamma D (\bp\times \hbe_\beta ) \partial_\beta \rho
  \nonumber  \\
	= &- \frac{D}{2A}  (\bp \times \bj) + \gamma D (\bp \times \nabla)\rho 
	\times \mb_0.
\end{align}
In derivation of \Eq{eqn:ap_extq_1} and \Eq{eqn:ap_dmtq},  we utilize following properties for circularly polarized spin wave $\mb'$, supposing that  $\bc=c_\theta\hbe_\theta+c_\phi\hbe_\phi$ is an arbitrary vector transverse to $\hbe_r\equiv\mb_0$, 
\begin{subequations}
\begin{align}
	&\Avg{ (\bc\cdot \mb')\mb'}\nonumber \\
	=& \Avg{(c_\theta m_\theta+c_\phi m_\phi)(m_\theta \hbe_\theta+  m_\phi \hbe_\phi)}\nonumber \\
	=& \Avg{m_\theta^2}c_\theta \hbe_\theta+ \Avg{m_\phi^2} c_\phi \hbe_\phi\nonumber \\
	=&  \rho \bc,
\end{align}
and similarly
\begin{align}
	&\Avg{ (\bc\cdot \mb')\partial_\beta\mb'}\nonumber \\
	\approx & \Avg{(c_\theta  m_\theta+c_\phi m_\phi)(\partial_\beta m_\theta \hbe_\theta+  \partial_\beta m_\phi \hbe_\phi)}\nonumber \\
	\approx &   \langle m_\theta \partial_\beta m_\phi \rangle c_\theta \hbe_\phi +\langle m_\phi \partial_\beta m_\theta \rangle c_\phi \hbe_\theta  \nonumber \\
	=&  \frac{j_\beta}{2\gamma A} (\bc\times \mb_0).
	\end{align}
\end{subequations}

Inserting \Eq{eqn:ap_extq} and \Eq{eqn:ap_dmtq} to \Eq{eqn:ap_torque_ADK},  we then obtain the full set of magnonic torques in \Eq{eqn:tau_sw} of the main text.
In above derivations, we use $\mb_0\times (\dots \times \mb_0)$ to extract the transverse components of magnonic torques with respect to texture magnetization $\mb_0$, to ensure the constant length of magnetization. 
In addition, several simplifications and approximations are utilized in above derivations, to organize the  magnonic torques in a compact and meaningful form.

\section{Magnonic forces}

\label{sec:ap_mag_forces}
The Thiele equation formulated in \Eq{eqn:force_tx_sw} can be rewritten in a force balance form,  $F^0_\mu+F_\mu=0$.
Here, $F^0_\mu$ is the internal force experienced by magnetic texture in the parametric space $\qty{X_\mu}$,
\begin{align}
    \label{eqn:force_tx}
     F^0_\mu=& F^0_{B}+ F^0_{E}+F^0_{\alpha} \nonumber \\
     =& B^0_{\mu\nu}  \dot{X}_\nu +E^0_\mu- \alpha \Lambda^0_{\mu\nu} \dot{X}_\nu,
\end{align}
and  $F_\mu$ is the external magnonic force transformed from magnonic torques \Eq{eqn:tau_sw},
\begin{align}
	\label{eqn:force_sw}
	F_\mu=&F_\ssf{STT}+F_\ssf{DM}+F_\ssf{GL}+F_\ssf{PL} \nonumber \\
	=  &-\int \qty[   (b^0_{\mu\beta}+ b^D_{ \mu \beta} )j_\beta
		+2   e^0_\mu  \rho -    \gamma A\lambda^0_{\mu\beta}  \partial_\beta \rho]d\cV.
\end{align}
The Amp\`ere force $F_\ssf{STT}$ (electrostatic force $F_\ssf{GL}$) in  \Eq{eqn:force_sw} 
and its dual partner $F^0_B$ ($F^0_E$) in \Eq{eqn:force_tx}  share the same field $b^0$ ($e^0$), and they together serve to transfer angular and linear momenta between spin wave and magnetic texture.
Meanwhile, the adhesion force $F_\ssf{PL}$ and viscous force $F_{\alpha}^0$ share the same adhesion field $\lambda^0$, and they mimic the sliding friction between two contacting surfaces. 
After above pairing of external and internal forces, the Amp\`ere force $F_\ssf{DM}$ imposed on field $b^D$ is left out as the only unpaired magnonic force, highlighting the unique role of DMI.
When $F_\ssf{DM}$ has non-zero overlap with $F_\ssf{STT}$ or $F_\ssf{GL}$,
it acts as an additional source for angular and linear momenta, i.e., momentum conservation is destroyed by DMI.

Specifically for a domain wall, by inserting \Eq{eqn:dmw_xphi} into \Eq{eqn:force_sw} in parametric space $\qty{X,\Phi}$, the magnonic forces $\bF\equiv (F_X,F_{\Phi})$ are explicitly given by
\begin{subequations}
	\label{eqn:force_sw_1d}
	\begin{align}
	\bF_\ssf{STT}=&\ 2\rho_0 v\qty(0,1) , \quad \bF_\ssf{DM}= -2\rho_0 v (-2\kappa\tilde{D}\sin\Phi,\tilde{D}\cos\Phi),\\
	\bF_\ssf{GL}=&\ 2\rho_0 (0, \frac{\pi \gamma  D }{W} \sin\Phi), \quad \bF_\ssf{PL}=-4\rho_0 (\frac{\kappa \gamma A  } {W},0),
\end{align}
\end{subequations}
where $\rho_0$ denotes the spin wave density at domain wall center, and $\tilde{D}=D/D_c$ is the normalized DM strength with $D_c=4\sqrt{AK}/\pi$ the critical DM strength. In responses to above magnonic forces, the internal forces are: 
\begin{subequations}
	\label{eqn:force_tx_1d}
\begin{align}
\bF_B^0 =&\ 2(-\dot{\Phi},\dot{X}), \\
\bF_E^0 =&\ (0, -\pi \gamma  D \sin\Phi/2), \\
\bF_\alpha^0 =&\ -2\alpha (\dot{X}/W,W\dot{\Phi} ),
\end{align}
\end{subequations}
where the  gyroscopic coefficients are  $B_{\Phi X}^0=-B_{X \Phi}^0=2$, and the viscous coefficients are $\Lambda^0_{XX}=2/W$ and $\Lambda^0_{\Phi\Phi}=2W$. 
Collecting all forces above, the domain wall dynamics in \Eq{eqn:dw_XP} is reproduced.

\begin{figure}[tb]
	\centering
	\includegraphics[width=0.48 \textwidth,trim=0 0 0 0,clip]{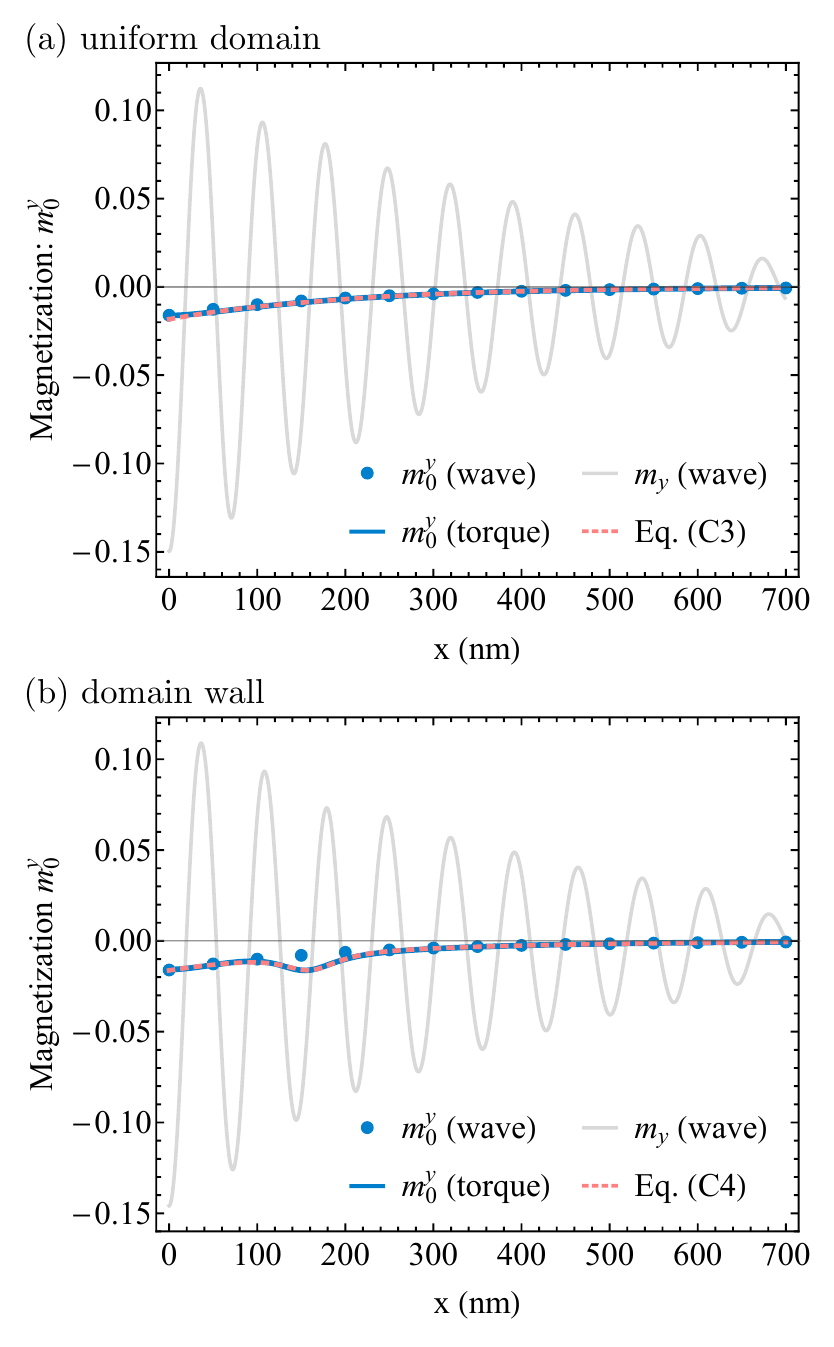}
	\caption{{\bf Modification of magnetic profile by magnonic DM torque in (a) a uniform domain and (b) a domain wall}. 
	The gray solid line is for total magnetization $m^y$ extracted from wave-based simulations, 
	and all other lines (dots) are for the static magnetization $m_0^y$. 
	The blue dots are from  wave-based simulations, the blue solid line is from torque-based simulations, and the red dashed line is from theoretical equations \eqref{eqn:ap_m0y_uni}\eqref{eqn:ap_m0y_dmw}.
	In all simulations and calculations, the DM strength is  $D=\SI{0.9e-3}{A}$, the magnetic damping is $\alpha=\SI{0.05}{}$.
	The spin wave density is  $\rho_0=\SI{9e-3}{}$ at the excitation point $x=\SI{0}{nm}$, and all magnetic data are taken after a relaxation time of $t=\SI{3}{ns}$.
	\label{fig:ap_tilt_dm}
	}
\end{figure}

\section{Asymmetric domain wall profile induced by Dzyaloshinkii-Moriya torque}
\label{sec:ap_mag_dm}

We first consider the case that the spin wave is travelling upon a uniform domain, in a magnetic wire along $x$-axis.
The overall magnetic field experienced by the uniform magnetization is 
\begin{align}
	\bh =  Km^z_0\hbz- \frac{D}{2\gamma A}\rho v \hby,
\end{align}
where the former is the anisotropic field, and the latter is effective field caused by magnonic DM torque according to \Eq{eqn:tau_sw_1d}.

At equilibrium, the magnetization aligns parallel to the magnetic field, $\mb_0\parallel \bh$, or $m_0^y/m_0^z=h_y/h_z$, therefore the magnetization is slightly titled out in $\hby$-direction with \cite{manchon_magnon-mediated_2014}
\begin{align}
	\label{eqn:ap_m0y_uni0}
	m_0^y\approx - \frac{D}{2\gamma AK}\rho v,
\end{align}
where only terms linear in $\rho$ are retained. 
It is noteworthy that to achieve the equilibrium magnetization state, finite magnetic damping ($\alpha\neq 0$) is required to release redundant energies.
Therefore, the spin wave density $\rho$ decays exponentially along $x$ direction, and \Eq{eqn:ap_m0y_uni0} is further given by 
\begin{align}
	\label{eqn:ap_m0y_uni}
	m_0^y\approx - \frac{D}{2\gamma AK}\rho_s v\exp[-2\kappa(x-x_s)],
\end{align}
where $\rho_s$ is the spin wave density at the excitation point $x_s$.
Such a magnetization tilting is verified by both wave-based and torque-based simulations in Fig. \ref{fig:ap_tilt_dm}(a).

We then move to the case that spin wave travelling upon a domain wall. 
Since the DM torque maintains the same form outside/inside domain wall, we expect that  the magnetization tilting in \Eq{eqn:ap_m0y_uni} valid for both domains at two sides naturally extends to the domain wall region. 
Hence, the full domain wall profile reads
\begin{align}
	\label{eqn:ap_m0y_dmw}                                     
		m_0^y\approx & \sech\qty(\frac{x-X}{W})\sin\Phi- \frac{D}{2\gamma AK}\rho_s v\exp[-2\kappa(x-x_s)],
\end{align}
where the $1$st term is the Walker profile defined upon central position $X$ and rotation angle $\Phi$ of the domain wall.
As a result of DM torque, the domain wall profile apparently becomes asymmetric with respect to the central position $X$.
In Fig. \ref{fig:ap_tilt_dm}(b), the extra magnetization tilting is confirmed by the torque-based simulation. 
However, the magnetization tilting is not discernable in wave-based simulation, possibly due to relatively large and complicated spin wave oscillations. Nevertheless, the magnetization titling in \Eq{eqn:ap_m0y_dmw} indicates that the domain wall dynamics induced by DM torque cannot be fully captured by \Eq{eqn:dmw_xphi} with only two collective coordinates $X$ and $\Phi$.

%\bibliography{refs1}

\end{document}